\documentclass[runningheads]{llncs}
\usepackage{graphicx,psfrag,epsfig,graphbox}
\usepackage{subfigure}
\usepackage{float} 
\usepackage{amsmath} 
\usepackage{amssymb} 
\usepackage{array}
\usepackage{soul}
\usepackage{color}
\usepackage[most]{tcolorbox}
\usepackage{subfig}
\usepackage{tabularx}
\usepackage{multirow}

\begin{document}
\title{Deep Local-Global Refinement Network for Stent Analysis in IVOCT Images}
\author{ 
Yuyu Guo\inst{1} \and
Lei Bi\inst{2} \and
Ashnil Kumar\inst{2} \and 
Yue Gao\inst{3} \and
Ruiyan Zhang\inst{3} \and
Dagan Feng\inst{2} \and
Qian Wang\inst{1} \and
Jinman Kim\inst{2}}

\authorrunning{Y. Guo et al.}
%
\institute{Institute for Medical Imaging Technology, School of Biomedical
Engineering, Shanghai Jiao Tong University, China \\
\email{wang.qian@sjtu.edu.cn}\and
School of Computer Science, University of Sydney, Australia \\
\email{jinman.kim@sydney.edu.au}\and
Ruijin Hospital, Shanghai Jiaotong University School of Medicine, China}

\maketitle              
\begin{abstract}
Implantation of stents into coronary arteries is a common treatment option for patients with cardiovascular disease. Assessment of safety and efficacy of the stent implantation occurs via manual visual inspection of the neointimal coverage  from intravascular optical coherence tomography (IVOCT) images. However, such manual assessment requires the detection of thousands of strut points within the stent. This is a challenging, tedious, and time-consuming task because the strut points usually appear as small, irregular shaped objects with inhomogeneous textures, and are often occluded by shadows, artifacts, and vessel walls. Conventional methods based on textures, edge detection, or simple classifiers for automated detection of strut points in IVOCT images have low recall and precision as they are, unable to adequately represent the visual features of the strut point for detection. In this study, we propose a local-global refinement network to integrate local-patch content with global content for strut points detection from IVOCT images. Our method densely detects the potential strut points in local image patches and then refines them according to global appearance constraints to reduce false positives. Our experimental results on a clinical dataset of 7,000 IVOCT images demonstrated that our method outperformed the state-of-the-art methods with a recall of 0.92 and precision of 0.91 for strut points detection.

\keywords{Convolutional Neural Network (CNN) \and Intravascular Optical Coherence Tomography (IVOCT) \and Stent Analysis}
\end{abstract}

\section{Introduction}

\begin{figure}[!htbp]
    \centering
    \includegraphics[width = 0.8 \textwidth]{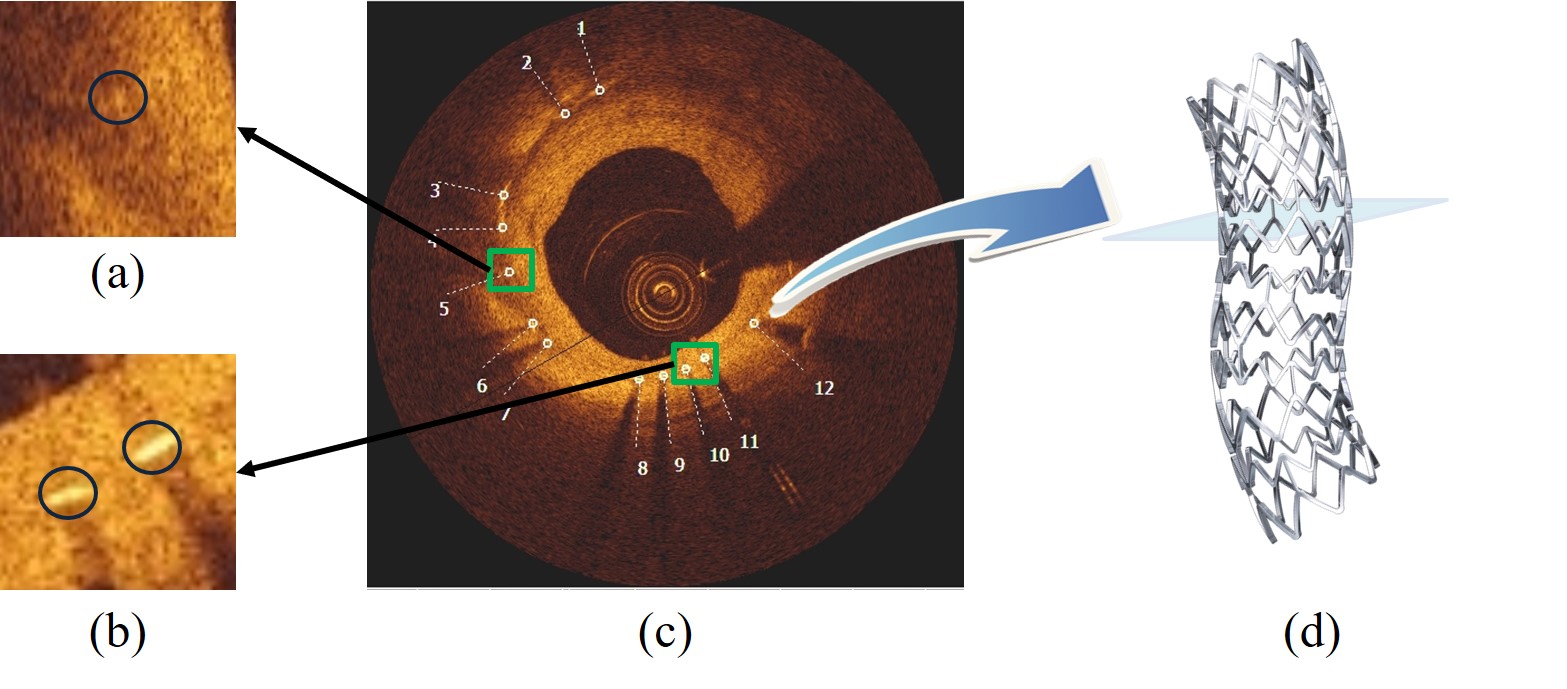}
    \caption{Common vascular features in IVOCT images. The middle image (c) shows the manually labeled stent struts (12 in total), with two green bounding boxes of struts (a) and (b).}
    \label{fig:1}
\end{figure}

Accurate assessment of neointimal coverage after stent implantation in intravascular optical coherence tomography (IVOCT) images is important to ensure the safety and efficacy of the Percutaneous Coronary Intervention procedure \cite{otsuka2015neoatherosclerosis}. Unfortunately, manual assessment requires the detection and analysis of thousands of struts within the stent, which is a challenging, tedious and time-consuming task. As shown in Fig. \ref{fig:1}, the stent struts are small, and the visual characteristics of the region covering the thick intima (innermost layer of the artery) may make the struts inconspicuous.

Motivated by these challenges, a number of automated detection methods have been proposed. Existing methods typically use handcrafted features to encode the candidate strut points and then apply supervised classification to identify the struts. Commonly used handcrafted features and supervised approaches include shadow feature \cite{ughi2012automatic,wang2013automatic}, decision trees \cite{lu2012automatic} and wavelet based detection \cite{mandelias2013automatic}. Besides, some studies used lumen segmentation \cite{nam2016automated,yong2017linear} and stent shape models \cite{ciompi2016computer} to constrain the search space for the potential struts candidates. However, all these methods rely on effective pre-processing steps, such as denoising, illumination corrections and detecting lumen boundaries for producing accurate results, which thereby restricting its generalizability. 

An alternative is to derive features using convolutional neural networks (CNNs) which have achieved great success in medical imaging analysis \cite{litjens2017survey}. The results from the use of CNN architectures including U-Net \cite{ronneberger2015u} and FCN \cite{long2015fully} demonstrate their performances in accurate detection and segmentation for large sized objects. However, because of the downsampling of the image to enlarge the receptive field and to encode the global information, the application of these CNNs to strut detection has not been validated.

We propose an automated method for stent detection in IVOCT images that overcomes the limitations above mentioned. We leverage CNNs for their ability to combine low-level appearance information with high-level semantic information in a hierarchical manner. A local network to densely detect potentially similar-struts in the image patches, and a global network that uses image appearance information to iteratively refines/removes the false predictions that are less likely to be struts. We have named our method the deep local-global refinement network (LGRN). We contribute the following to the state-of-the-art:
\begin{itemize}
\item[$\bullet$] To the best of our knowledge, this is the first deep learning method for strut points detection. Our method removes the reliance on pre-processing steps such as denoising and illumination corrections.
\item[$\bullet$] Our coupling of a local network that has high recall with a global network that provides image level refinement, enables false positive reductions while maintaining high sensitivity and efficiency to strut point detection.
\item[$\bullet$] Our global network uses an appearance constrained attention module for false positive reduction, which preserves the detected struts that fit the overall appearance of the image.
\end{itemize}

\section{Method}
\subsection{Materials}
Our dataset consists of 57 patients with stent implanted for more than 1 year. Each patient has an average of about 127 IVOCT images with stent. All the IVOCT images were acquired using a C7-XR OCT system (St. Jude Medical, St. Paul, MN, USA) with a 2.7f Dragonfly imaging catheter. Each IVOCT image has a resolution of $714\times714$ pixels. A cardiologist performed manual annotation of the struts and lumen on all the IVOCT images. And we follow the same protocol as in \cite{merget2018robust} to design a morphological filter applied to the annotated struts to enlarge the size, which was then used as the ground truth for evaluation.

\begin{figure}[!htbp]
\centering\includegraphics[width=1 \textwidth]{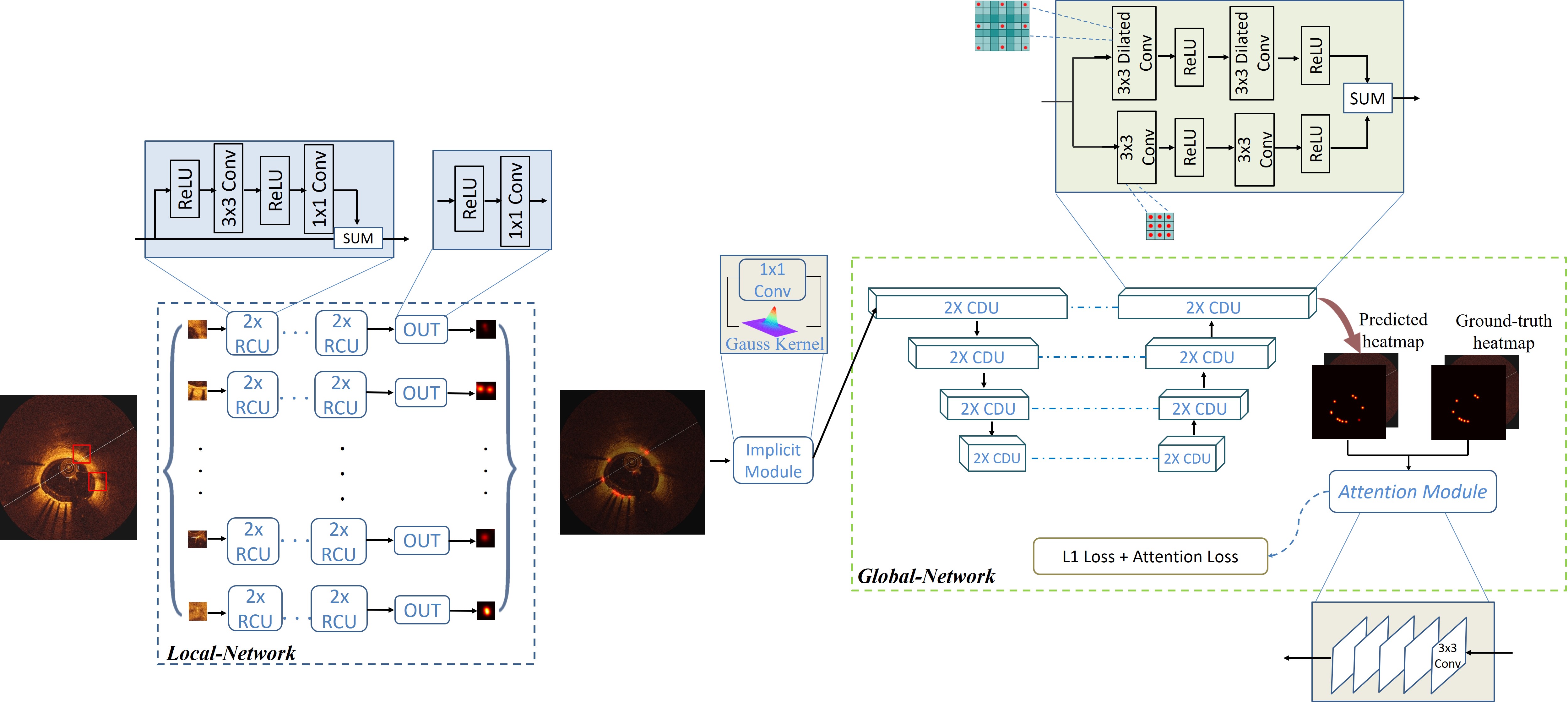}
\caption{Overall structure of the LGRN.}
\label{fig:2}
\end{figure}

\subsection{Deep Local-Global Refinement Network} 

Fig. \ref{fig:2} shows the overview of the proposed local-global refinement network. Initially, Local-Network is applied to the input IVOCT image to detect all the potential struts via small input patches. After that, the detected struts together with the original IVOCT image are used as the input to Global-Network for refinement, where an appearance constrained attention module is applied to guide the overall spatial distribution of all the struts and to remove all falsely detected struts.

\subsubsection{Local-Network:} A patch-based deep CNN is used as the Local-Network for detecting all the strut points. It consists of 9 zero-padded convolutional layers with kernel size of 3 and stride of 1. Residual block is used to connect each adjacent layer. At the end of the Local-Network we use a linear $1\times1$ convolution layer and a Gaussian convolution kernel, where the $1\times1$ convolution is used to compensate for batch normalization, and the Gaussian convolution kernel is used to smooth the output, e.g., suppressing the path artifacts of the Local-Network. The patch size is set to $64\times64$ for the Local-Network. The network is trained with \textit{$L1$} Loss. 

\subsubsection{Global-Network:} The purpose of the Global-Network is to extract high-level semantic information that can be used to guide the refinement of all the detected struts. The Global-Network uses a modified U-net \cite{ronneberger2015u} architecture. There are 4 downsampling layers, each with a \textit{$2 \times 2$} max-pooling operator, and 4 upsampling layers. At each down/up-sampling layer, we repeat the following parallel architecture module: one sub-branch with \textit{$3 \times 3$} convolutional kernel, and another sub-branch with \textit{$3 \times 3$} dilated convolution and 2 dilations. The outputs of these two sub-branches are added at the end of this module. This combination of regular convolution and dilated convolution has a larger receptive field thus it can learn visual characteristics that assist in inferring struts with less visual features. Therefore, global context constraints can ensure the accuracy of the prediction results, while local context learning can improve the sensitivity of the model to detect the object. However, the uneven distribution of the background and foreground makes it difficult for the global network to converge during training. To overcome this, we added an appearance constrained attention module to guide its convergence, where we used another 5 layers CNN network to different whether the predicted detection results have the similar appearance to the ground truth. To facilitate the learning process, we used two loss functions (\textit{$\ell_{similar}$} and \textit{$\ell_{attention}$}) as:

 \begin{equation}\label{eq:3}
    \ell_{similar} = \ell_{L1}\Big(\lceil{M}\rceil*P, M\Big) + \ell_{L1}\bigg(\Big(1-\lceil{M}\rceil\Big)*P, 0\bigg )
\end{equation}

where \textit{$\lceil{M}\rceil$} is the ground truth annotation. \textit{P} indicates the predicted results. It's used to balance the uneven distribution of foreground and background. We also add an attention loss to constrain the overall appearance of the predicted struts and this can be defined as:

\begin{equation}\label{eq:5}
    \ell_{attention} = log\Big(A(M,I)\Big) + log\Big(1 - A(P,I)\Big)
\end{equation}

where \textit{$A(\cdot)$} indicates the attention module that discriminates whether \textit{$P$} is similar appearance to the ground truth.

\subsection{Implementation Details}
We pre-processed the dataset with maximum normalization and cropped all images to \textit{$512 \times 512$}. Both Local-Net and Global-Network were trained for 80 epochs with an Adam optimizer at an initial learning rate of 0.001 and batch size of 1. It took an average of 15 hours to train on an 11GB Nvidia 1080Ti GPU.

\section{Results and Discussions}

\subsection{Experimental Setup}
We randomly divided the dataset into a training set (30 patients, 3873 images) and a test set (27 patients, 3352 images) for evaluation. 
We performed the following experiments on the dataset: (a) comparison of the performance of our method with the state-of-the-art methods; and (b) analysis of the performance of each component in our method.
The state-of-the-art methods include: (i) Wang, Ancong et al. \cite{wang20153} - Bayesian network based detection method; (ii) Faster-RCNN \cite{ren2015faster} - Region Proposal Network (RPN) that is trained end-to-end to generate high-quality region proposals for detection. (iii) Lu, Hong et. al \cite{lu2012automatic} - Bagged decision trees classifier for classifying candidate struts using structure features, and (iv) Nam, Hyeong Soo et. al. \cite{nam2016automated} - Neural network classifier to classify the features from gradient images. For \cite{wang20153}, due to the unavailability of the algorithm source code, we refer to the published result as a reference acknowledging that the dataset is different. We use recall and precision for the evaluation (we followed \cite{wang20153} to define the true positive detection if they are within 5 pixels of the ground-truth).

\subsection{State-of-the-art methods comparison}

Table~\ref{table:1} shows the detection results of our method compared to the state-of-the-art methods, where it increases recall by 1.2\% and precision by 4.7\%, relative to the second best results from Faster-RCNN,  as shown in Fig 3(a). 

\begin{table}[h]
\centering
\caption{Detection results compared to the state-of-the-art methods}\label{tab1}
\begin{tabular}{| p{3cm}<{\centering} | p{3cm}<{\centering} | p{3cm}<{\centering} |} 
\hline
Method &  Recall & Precision\\
\hline
Hyeong et al. \cite{nam2016automated} &  0.894 & 0.818\\
Hong et al. \cite{lu2012automatic} &  0.804 & 0.826\\
Ancong et al. \cite{wang20153}& 0.910 & 0.840\\
Faster-RCNN \cite{ren2015faster}&  0.913 & 0.856\\
\hline

Local-Network&  0.914 & 0.833\\
Global-Network&  0.903 & 0.876\\
Ours& {\bfseries 0.925} & {\bfseries 0.903}\\
\hline
\end{tabular}
\label{table:1}
\end{table}

\begin{figure}[h]
\centering
\subfigure[]{
\vspace{0pt}
\includegraphics[align=t,width=0.43 \textwidth]{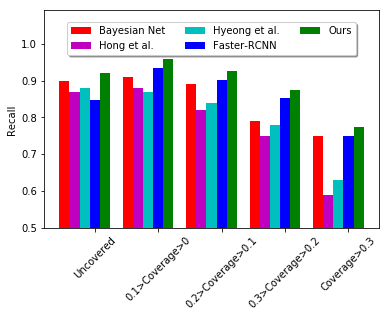}
}
\subfigure[]{
\includegraphics[align=t,width=0.43 \textwidth]{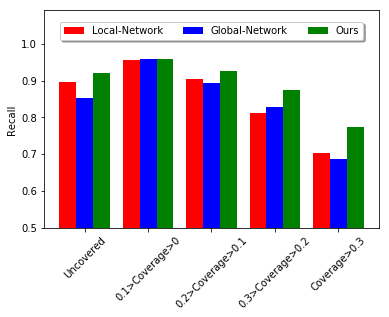} 
} 
\caption{Strut detection results: (a) Comparison of our method to the existing methods, (b) component analysis of our method.}
\label{fig:3}
\end{figure}

\subsection{Component Analysis}
Table \ref{table:1} and Fig. 3(b) show the detection results of our method at individual stages. Fig. 5(a) shows the two example detection results with various thickness coverage. The Local-Network shows the higher recall while Global-Network achieved better precision results (as shown in Fig. 5(c) and Fig. 5(d)). As exemplified in Fig. 5(e), the proposed method integrated both Local Network and Global-Network and achieved a better consistent performance in recall and precision.

\subsection{Discussion}

\begin{figure}[h]
\centering
\subfigure[Image]{
\begin{minipage}[]{0.15\textwidth}
\centering
\includegraphics[width=1 \textwidth]{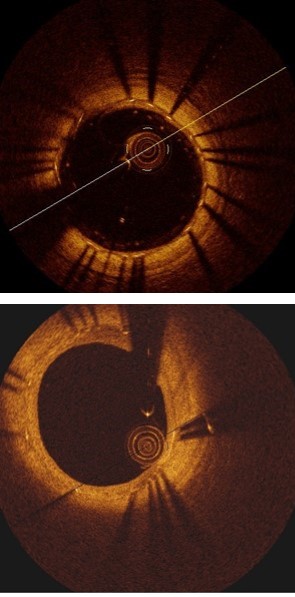}
\end{minipage}
}%
\subfigure[Labels]{
\begin{minipage}[]{0.15\textwidth}
\centering
\includegraphics[width=1 \textwidth]{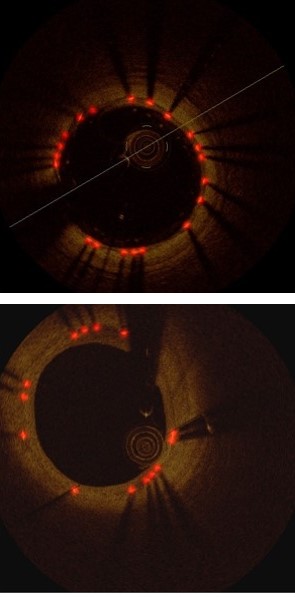}
\end{minipage}
}%
\subfigure[F-RCNN]{
\begin{minipage}[]{0.15\textwidth}
\centering
\includegraphics[width=1 \textwidth]{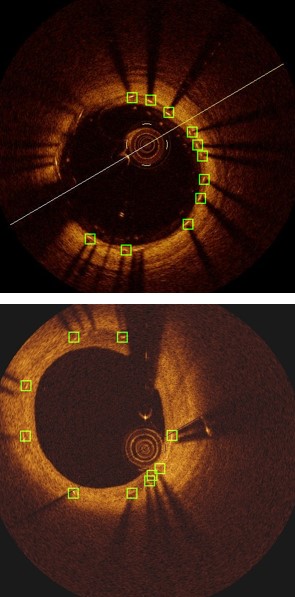}
\end{minipage}
}%
\subfigure[Hong.]{
\begin{minipage}[]{0.15\textwidth}
\centering
\includegraphics[width=1 \textwidth]{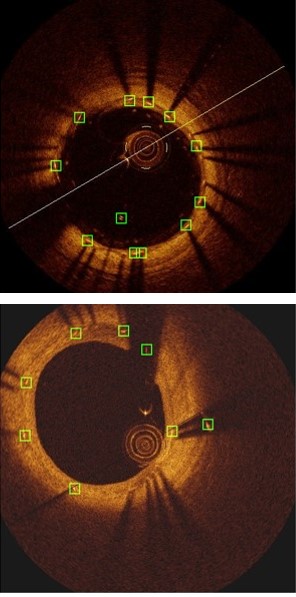}
\end{minipage}
}%
\subfigure[Hyeong.]{
\begin{minipage}[]{0.15\textwidth}
\centering
\includegraphics[width=1 \textwidth]{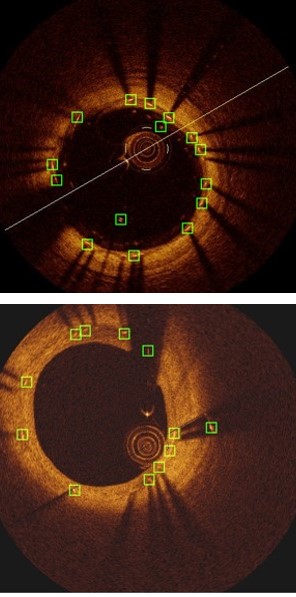}
\end{minipage}
}%
\subfigure[Ours]{
\begin{minipage}[]{0.15\textwidth}
\centering
\includegraphics[width=1 \textwidth]{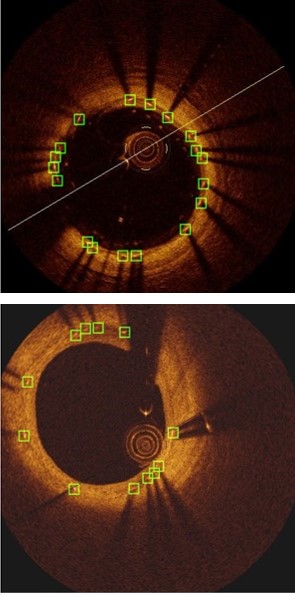}
\end{minipage}
}
\centering
\caption{Comparison of our detection results with existing comparison methods}
\label{fig:4}
\end{figure}

Table 1 and Fig. 3(a) illustrate our method achieved the overall best performance when compared to the existing methods for strut detection. The traditional methods (Hyeong et al.\cite{nam2016automated}, Hong et al.\cite{lu2012automatic} and Ancong et al.\cite{wang20153}) using hand-crafted features with conventional classifiers achieved competitive performance when compared with Faster-RCNN method. Fig. 4(d) and Fig. 4(e) show two example results where both Hyeong et al and Hong et al methods fail to detect strut points where there is low-contrast to the background. In contrast, Faster-RCNN has the ability to combine deep semantic information and shallow appearance information in a hierarchical manner that enables it to encode image-wide location information and semantic characteristics. However, Faster-RCNN lacks constrain of the overall appearance of all the struts. Consequently, Faster-RCNN generates poor detection results for small struts (as shown in Fig. 4(c)).

\begin{figure}[h]
\centering
\subfigure[Image]{
\begin{minipage}[]{0.15\textwidth}
\centering
\includegraphics[width=1 \textwidth]{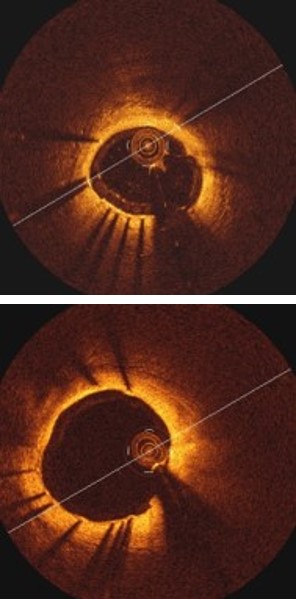}  
\end{minipage}
}%
\subfigure[Label]{
\begin{minipage}[]{0.15\textwidth}
\centering
\includegraphics[width=1 \textwidth]{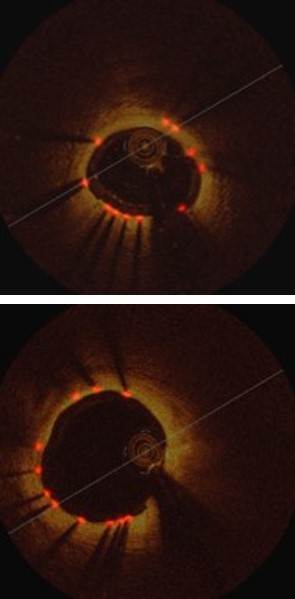}   
\end{minipage}
}%
\subfigure[Locals.]{
\begin{minipage}[]{0.15\textwidth}
\centering
\includegraphics[width=1 \textwidth]{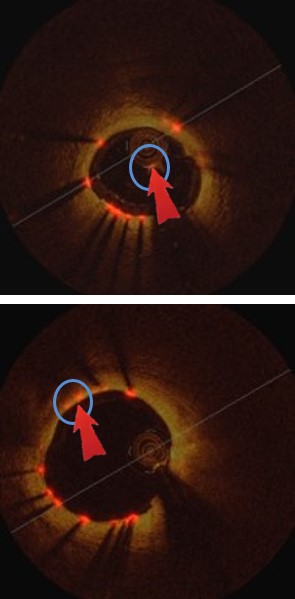} 
\end{minipage}
}%
\subfigure[Global.]{
\begin{minipage}[]{0.15\textwidth}
\centering
\includegraphics[width=1 \textwidth]{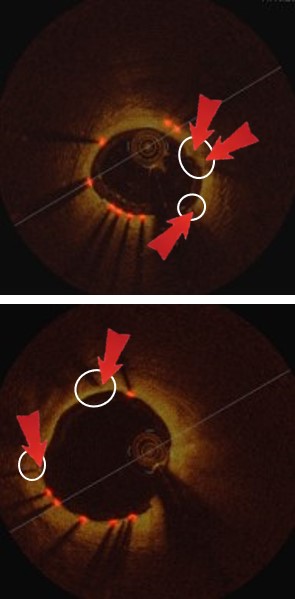} 
\end{minipage}
}%
\subfigure[Ours]{
\begin{minipage}[]{0.15\textwidth}
\centering
\includegraphics[width=1 \textwidth]{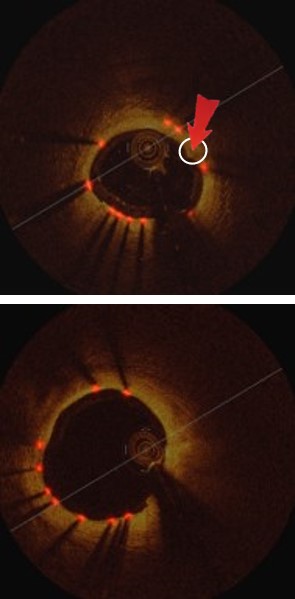}  
\end{minipage}
}
\centering
\caption{Example results of struts detection. Red arrows indicate the detection errors.}
\label{fig:5}
\end{figure}

Table 1, Fig. 3(b) and Fig. 5 compared the main components of our method individually to quantify their contributions to the final detection results. These results demonstrate that Local-Network has higher recall and we attribute this to the usage of patch-based network to detect all the potential strut candidates. In contrast, Global-Network achieved higher precision for its ability by adding global context, e.g., appearance information, as part of the learning process, which ensures all the detected struts are consistent with the shape of the stent. Table 1, Fig. 3(b) and Fig. 5 also show the advantages from our combination which integrates complementary detection results produced at individual components.

\section{Conclusion}
We propose a deep learning-based method for stent struts detection for IVOCT images. We achieved state-of-the-art struts detection performance via a local-global refinement network, where we detected potential struts which were then refined according to global appearance constraints to reduce false positives. Our experimental results demonstrate that our method achieved higher accuracy when compared to the existing state-of-the-art methods on a large clinical dataset.\\

\noindent\textbf{Acknowledgement} This work was supported in part by Australia Research Council (ARC) grants (LP140100686 and IC170100022), the University of Sydney – Shanghai Jiao Tong University Joint Research Alliance (USYD-SJTU JRA) grants and STCSM grant (17411953300).

%
%
%
%
%
%
\bibliographystyle{splncs04}
\bibliography{oct_ref}

\end{document}